\documentclass[review]{elsarticle}

\usepackage{lineno,hyperref}
\modulolinenumbers[5]

\journal{Optics Communications}

\newcommand{\sech}{\mbox{sech}}

\begin{document}

\begin{frontmatter}

\title{  Optical solitons in periodically managed PT-symmetric media}

\author{F. Kh. Abdullaev, R. M. Galimzyanov }
\address{Physical-Technical Institute, Uzbekistan Academy of Sciences, Tashkent, Uzbekistan}




\begin{abstract}
The dynamics of light beams in the nonlinear optical media with periodically modulated in the longtidunal direction parity-time distribution of the complex refractive index
is investigated. The possibility of dynamical stabilization of PT-symmetric solitons is demonstrated.
\end{abstract}

\begin{keyword}
parity time symmetry \sep optical soliton \sep periodic management \sep imaginary Kapitsa problem \sep dynamical stabilization
\end{keyword}

\end{frontmatter}

\linenumbers


\section{Introduction}

Investigation of nonlinear optical wave phenomena in media with
the parity-time symmetry properties of the complex spatially
modulated refraction index attracts a great attention last time
\cite{KYZ,Dmitriev}. Mainly it is related with the important
properties of the systems with non-Hermitian hamiltonian with
PT-symmetry. Such systems for values of the gain/loss parameter
lower than critical values have real eigenvalues \cite{Bender}.
This phenomenon has direct analogies in the optical systems. For example in the paraxial
approximation the wave equation for electromagnetic waves has the form of the Schr\"odinger equation
 with
the spatial distribution of the gain/loss parameters. The PT
symmetry constrain  leads to the existence of the stationary modes
in the non-conservative optical systems. In nonlinear optical
media described by the PT-symmetric extended nonlinear
Schr\"odinger equation  it leads to the existence of families of
optical solitons \cite{Musslim,AKKZ}.
Now a great interest has been attracted to management of solitary
waves in media with parity-time invariant properties. In
particular the periodically modulated optical systems with
balanced gain and loss has been studied in \cite{Kiv1}, where in
the high frequency limit an effective PT-symmetric system has been
obtained from initially non - PT symmetric optical system. Note
the periodic variation in the longitudinal direction (LD)
involving the real part of the refractive index was considered
there. Another possibility is to consider periodic variations in
the longitudinal direction, involving also the imaginary part of
the refractive index, i.e. the imaginary part of the potential
$V_I$ in the corresponding Schr\"odinger equation. These
possibilities are investigated in the nonlinear directional
coupler with periodically modulated gain/loss along LD
\cite{Mal1,Abd1}. The stabilization of nonlinear modes has been
found. The considered problem can be called as {\it  the imaginary
Kapitza problem}. In the work \cite{Longhi} a  linear
Schr\"odinger equation with oscillating imaginary Gaussian
potential has been investigated. It was shown that in the HFL the
spectrum of quasi-energy is real valued leading to the existence
of stationary modes. The results have applications to the
stability of optical resonators with variable reflectivity.
Another example is the BEC with spatially distributed gain//loss
parameters, described by the Gross-Pitevskii equation with PT
symmetric potential \cite{CW}.
 It
is of interest to study the imaginary Kapitza problem for the case
of {\it nonlinear optical media}.
 We will consider dynamics
of solitons in media with the Kerr nonlinearity with rapidly managed PT-symmetric  potential.
In particular we will study the possibility of stabilization of unstable nonlinear modes.

The paper is organized as follows: In section 2 we formulate the
model and describe the procedure of obtaining averaged equation
for the GP equation with a PT-potential being periodically varied
in the longthidunal direction  and inhomogeneous in space. The
results of numerical simulations of the full and averaged NLS
equations for optical soliton propagation are presented in section
3.

\section{The model: Averaged NLS  equation}
Let us consider propagation of a light in the nonlinear media with
modulated in the space complex refraction index. In the parabolic
approximation the electric field propagation is described by the
wave equation \cite{AK}:
\begin{equation}
2ik_0 n_0 E_Z + E_{xx} +2k_0^2 n_0 \Delta n(x,z)E + 2k_0^2 n_0 n_2 |E|^2 E = 0,
\end{equation}
where $ k_0=2\pi/\lambda$, $\Delta n(x,z)=n_R(x)+ i n_I(x,z)$  is the complex
refraction index. Introducing dimensionless variables
$$
t=Z/L_d, x =X/w_0, u=(k_0|n_2|L_d)^{1/2}E,\  U(x,z)=-k_0 L_d \Delta n(x,z),
$$
where $L_d=k_0 n_0 w_0^2$ is the diffraction length, $w_0$ is the
beam initial width , we obtain the following wave  equation:
\begin{equation}\label{nls}
iu_t + u_{xx}-V(x,t)u + 2|u|^2 u=0.
\end{equation}
Analogous equation also describes dynamics of BEC with attractive
interaction in the complex trap potential, namely, the time
dependent GP equation
\begin{equation}
i\hbar\psi_{\tau} = -\frac{\hbar^2}{2m}\Delta\psi + V_{tr}(r,t)\psi +
g|\psi|^2 \psi, \label{genGP}
\end{equation}
where $m$ is the atom mass, $g = 4\pi\hbar^2 a_{s}/m$ and  $a_{s}$
is the atomic scattering length. $a_{s}>0$ corresponds to the BEC
with repulsive interaction between atoms and $a_{s} < 0$ to the
attractive interaction. We restrict ourselves by considering a
cigar-shaped condensate that allows us to deal with a quasi 1D
mathematical model (\ref{nls}). The  complex trap potential in BEC
has been realized recently in \cite{Oberthaler}.

We will consider next model of time modulated potential.

$$V(x,t) = \alpha(t)V_{PT}(x).$$

The another possible model is with modulation in time of the imaginary part
of the potential
$$
V(x,t)=V_R(x) + i\alpha(t)V_I(x),
$$
can also be described by considering in this work method.
We come to the following governing equation
\begin{equation}\label{govGP}
iu_t +   u_{xx} - \alpha(t)V_{PT}(x) u + 2|u|^2 u = 0 ,
\end{equation}
where $V_{PT}(x)$ is a PT- symmetric  potential in $X,t$-space,
$\alpha(t)$ describes the time modulations of the potential.
$V_{PT}(x)=f(x)$ is  a complex function with even real and odd
imaginary parts, $\alpha(t) = \alpha_0 + \gamma(t) $. The
modulation of potential  is supposed to be periodic in time
\begin{equation}\label{osc}
\gamma(t+T) = \gamma(t), \ \ \  \int_0^T
\gamma(t^{\prime})dt^{\prime} = 0.
\end{equation}
Temporal  modulation is chosen as
\begin{equation}\label{gam}
\gamma (t) = \alpha_1 \cos(\Omega t).
\end{equation}
We suppose the modulation frequency $\Omega$ to be large ($\Omega
\gg 1$). Thus in our problem  a value $\epsilon = \alpha_1 / \Omega$
emerges which is a characteristic of the potential strength modulation and the frequency of  rapid
oscillations.

\subsection{Small strength of the trap perturbation $\epsilon = O(0)$}

The field $u(x,t)$ can be represented in the form of sum of slowly
and rapidly varying parts $U(x,t)$ and $\xi(x,t)$
\begin{equation}\label{U}
u(x,t) = U(x,t) + \xi(x,t) .
\end{equation}

For obtaining equation for the averaged  field $u(x,t)$ we will
present the rapidly varying part of the field as a Fourier series
expansion \cite{Kivshar1995,AbdGal}:
\begin{equation}\label{xi}
\xi = A\cos(\Omega t) + B\sin(\Omega t) + C\cos(2\Omega t) +
D\sin(2\Omega t)+ ... \ ,
\end{equation}
where $A, B, C$ and $D$ are slowly varying in time small functions
of variables $x$ and $t$ of the order $O(1)$. By substituting
equations (\ref{U}) and (\ref{xi}) into (\ref{govGP}) we obtain
the next set of equations for the slowly varying field $U$ and
coefficients of the expansion of the rapidly varying component

\begin{eqnarray}\label{eq1}
iU_t +  U_{xx} + 2|U|^2 U + 2U(|A|^2 + |B|^2 + |C|^2 + |D|^2) + \nonumber\\
U^{*}(A^2 + B^2 + C^2 + D^2 ) + A^{*}BD + AB^{*}D + ABD^{*} + \nonumber\\
C(|A|^2 - |B|^2) + \frac{1}{2}C^{*}(A^2 - B^2) = \alpha_0 f(x) U +
\frac{\alpha_1}{2} f(x) A , \end{eqnarray}

\begin{eqnarray}\label{eq2}
iA_t + i\Omega B +  A_{xx}  + 4|U|^2 A + 2U(AC^{*} + A^{*}C +
BD^{*} + B^{*}D) +  \nonumber \\
2U^{*}(AC + BD) + 2U^2 A^{*} + A \left( \frac{3}{2}|A|^2 + |B|^2 +
2|C|^2 + 2|D|^2 \right) + \nonumber \\
A^{*} \left( \frac{1}{2}|B|^2 + |C|^2 + |D|^2 \right) = \alpha_0
f(x) A + \frac{\alpha_1}{2} f(x) C + \alpha_1 f(x) U,
\end{eqnarray}

\begin{eqnarray}\label{eq3}
iB_t - i\Omega A +  B_{xx}  + 4|U|^2 B + 2U(AD^{*} + A^{*}D -
BC^{*} - B^{*}C) +  \nonumber \\
2U^{*}(AD - BC) + 2U^2 B^{*} + B \left(|A|^2 + \frac{3}{2}|B|^2 +
2|C|^2 + 2|D|^2 \right) + \nonumber \\
B^{*} \left( \frac{1}{2}|A|^2 + |C|^2 + |D|^2 \right) = \alpha_0
f(x) B + \alpha_1 f(x) D,
\end{eqnarray}

\begin{eqnarray}\label{eq4}
iC_t + 2i\Omega D +  C_{xx}  + 4|U|^2 C + 2U(|A|^2 - |B|^2) +
\nonumber \\
U^{*}(A^2 - B^2) + 2U^2 C^{*} + C \left( 2|A|^2 + 2|B|^2 + \frac{3}{2}|C|^2 + |D|^2 \right) + \nonumber \\
C^{*} \left( |A|^2 + |B|^2 + \frac{1}{2}|D|^2 \right) = \alpha_0
f(x) C + \frac{\alpha_1}{2} f(x) A ,
\end{eqnarray}

\begin{eqnarray}\label{eq5}
iD_t - 2i\Omega C +  D_{xx} + 4|U|^2 D + 2U(AB^{*} + A^{*}B)+
\nonumber \\
2U^{*} A B + 2U^2 D^{*} + D \left( 2|A|^2 + 2|B|^2 + |C|^2 + \frac{3}{2}||D|^2 \right) + \nonumber \\
D^{*} \left( A^2 + B^2 + \frac{1}{2}C^2 \right) = \alpha_0 f(x) D
+ \frac{\alpha_1}{2} f(x) B .
\end{eqnarray}

The parameter $\Omega$ is supposed to be  $\gg 1$. Then we can
introduce a small parameter $\epsilon=1/\Omega$. Considering
Eq.~(\ref{eq2}) and neglecting all small terms of the order of
$O(\epsilon^2)$ we come to $$iB + O(\epsilon^2) =
\frac{\alpha_1}{\Omega}f(x)U + O(\epsilon^2).$$ So we have
solution for unknown $B$
\begin{equation}\label{B}
B = -i\epsilon \alpha_1 f(x)U.
\end{equation}
Thus structure of equations (\ref{eq2} - \ref{eq5}) allows to
suppose that $$B\approx \epsilon, \ \ A\approx \epsilon^2, \ \
C\approx \epsilon^2, \ \ D\approx \epsilon^3 .$$ The rest of
parameters $A$ and $C$ are
\begin{eqnarray}\label{ADC}
A = -i\epsilon^2 \alpha_1 f(x) U_t - \epsilon^2  \alpha_1
(f(x)U)_{xx} - 2\epsilon^2 \alpha_1 f(x)|U|^2 U  + \epsilon^2
\alpha_0 \alpha_1 f^2(x) U, \nonumber \\
C = \epsilon^2  {\alpha_1}^2\frac{f^2(x)}{4} U.
\end{eqnarray}
Let's consider Eq.~(\ref{eq1}). Holding only terms  of order
$\epsilon^2 $ the equation becomes
\begin{eqnarray}\label{eq1s}
iU_t +  U_{xx} + 2 |U|^2 U + 2U|B|^2 + U^{*}B^2 = \alpha_0
f(x)U + \frac{\alpha_1}{2}f(x) A .
\end{eqnarray}
Substituting expressions for $A,B,C$ from (\ref{B}), (\ref{ADC})
into Eq.~(\ref{eq1s}), we obtain an averaged equation for $U$
\begin{eqnarray}\label{slowU}
iU_t +  U_{xx} + 2 |U|^2 U + 2\epsilon^2 {\alpha_1}^2
|f(x)|^2 |U|^2 U - \epsilon^2 {\alpha_1}^2 f^2 (x) |U|^2 U =
\nonumber \\
\alpha_0 f(x)U - i\frac{\epsilon^2}{2} {\alpha_1}^2 f(x)U_{t} -
 \frac{\epsilon^2}{2}{\alpha_1}^2 f(x) (f(x)U)_{xx} -
\nonumber \\
\epsilon^2 {\alpha_1}^2 [2f(x) - f^*(x)] f(x) |U|^2 U +
\frac{\epsilon^2}{2} {\alpha_1}^2 \alpha_0 f^3(x) U.
\end{eqnarray}
Hereafter we redesignate  $\epsilon = \alpha_1 /\Omega$. Then
introducing new field $V(x,t)$
\begin{equation}\label{V}
\tilde{U}(x,t) = \left( 1 + \frac{\epsilon^2}{2} f^2(x)
\right)^{1/2}U(x,t),
\end{equation}
into Eq.~(\ref{slowU}) we have finally the following equation
\begin{equation}\label{avGP}
i\tilde{U}_t +  \tilde{U}_{xx} + 2 (1+ G_{eff})
|\tilde{U}|^2 \tilde{U} = W_{eff}(x)\tilde{U}.
\end{equation}
 The averaged equation is a modified NLS equation with
 an effective slowly varying linear and nonlinear  potentials
\begin{equation}\label{Veff}
 \ W_{eff}(x) = \alpha_{0} f(x) + \frac{\epsilon^2}{2}
f^2_{x}(x), \  2G_{eff}|\psi|^2 = 2\left( 1 + \epsilon^2(Imf(x))^2 \right)|\tilde{U}|^2.
\end{equation}

\subsection{Strong management}

Let us consider for simplicity  the case of pure imaginary
potential ($V_R=0$). To investigate the dynamics of beam in the
case of strong management  it is useful to perform transformation
of the field
\begin{equation}\label{eq2f}
u(x,t) = U(x,t)e^{-\Gamma(t)V_I(x)},
\end{equation}
where $\Gamma(t)$ is the antiderivative for $\gamma(t)$
$$
\Gamma(t)=\int_0^t \gamma(t^{\prime})dt^{\prime}.
$$
Here we assume that field  $U(x,t)$ is slowly varying on the
period of oscillations of complex potential. We will check the
consistency of this assumption s by the multiscale perturbation
theory below. Substituting this expression into Eq.~(\ref{eq1}) we
obtain the equation for $U(x,t)$:
\begin{equation}
iU_t +  U_{xx} -2U_x\Gamma V_{I,x} -U\Gamma V_{I,xx} + U \Gamma^2
V_{I,x}^2 + 2|U|^2 U e^{-2\Gamma V_I(x)}=0
\end{equation}
Let us consider the case  of periodic modulations
$$
\gamma(t) =\gamma_0 \cos(\omega t).
$$
and {\it the management is strong} i.e. $\gamma_0, \omega \gg 1, \gamma_0/\omega =O(1)$. Averaging this equation over period of modulations we have averaged equation
\begin{equation}\label{lstavGP}
iU_t + U_{xx} +  \frac{\gamma_0^2}{2\Omega^2}V_{I,x}^2 U +
2I_0(\frac{2\gamma_0}{\Omega} V_I(x))|U|^2 U  =0.
\end{equation}
Here we take $ <\Gamma> =0, \ <\Gamma^2> = \gamma_0^2/2\omega^2, \
<...> = \frac{1}{T}\int_0^T (...)dt, \ T = 2\pi/\omega $. $I_0(x)$
is the modified Bessel function of the first kind
$$
I_0 (x)= \frac{1}{2\pi}\int_0^{2\pi}e^{x\sin( t)}dt.
$$

From this averaged equation we can now conclude that dynamics is
described now by {\it conservative } modified NLS equation  with
the effective linear potential
$$
-  \frac{\gamma_0^2}{2\Omega^2}V_{I,x}^2
$$
and the effective nonlinear potential induced by spatially varying Kerr nonlinearity
$$
I_0[\frac{2\gamma_0}{\Omega} V_I(x)] >1.
$$

Considering the case of weak management when $2\gamma_0/\Omega \ll
1$ and expanding in Eq.~(\ref{lstavGP}) coefficients in power
series we obtain the averaged equation Eq.~(\ref{avGP}) for the
weak management case.

\section{Numerical results}
In numerical simulations we use the following PT potential which
was used in the works \cite{Ahmed}
\begin{eqnarray}\label{PTpot}
V_{PT}(x) = -V_0 \sech^2(x) - W_0 \sinh(x)\sech^2(x) ,
\end{eqnarray}
Soliton solution to the standard Gross-Pitaevskii equation
\begin{equation}\label{stGP}
iu_t +   u_{xx} - V_{PT}(x) u + 2|u|^2 u = 0 ,
\end{equation}
with potential (\ref{PTpot}) has the form \cite{Musslim}
\begin{equation}\label{solMus}
u_s(x,t) = \frac{u_0}{\sqrt{2}}\sech(x)\exp(i \mu
\arctan(\sinh(x))\exp(i\lambda t) ,
\end{equation}
where the solution parameters $\lambda = 1,  \ \ \ \mu = W_0 / 3 $
and $\ \  u_0 = \left( 2 - V_0 + \mu^2 \right)^{1/2}.$ As known
from the work \cite{Ahmed} PT-potential Eq.~(\ref{PTpot}) in
linear problem exhibits real spectrum provided that $W_0 \leq V_0
+ 1/4$. In other words it means that otherwise the potential leads
to unstable solutions of equation ~(\ref{stGP}).

\begin{figure}[htb]
\includegraphics[width=6cm,height=6cm,angle=-90,clip]{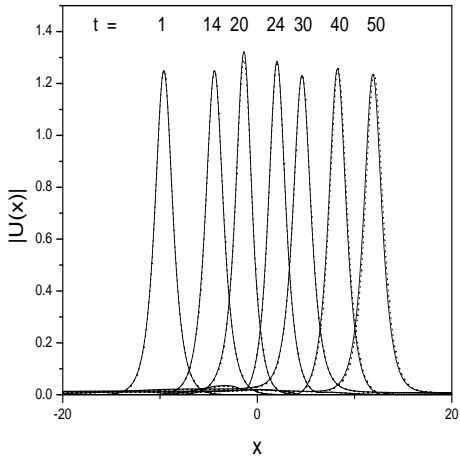}\quad
\caption{Evolution of the NLS soliton profile when travelling
through the imaginary PT-potential.} \label{Fig1}
\end{figure}

\begin{figure}[htb]
\includegraphics[width=6cm,height=6cm,angle=-90,clip]{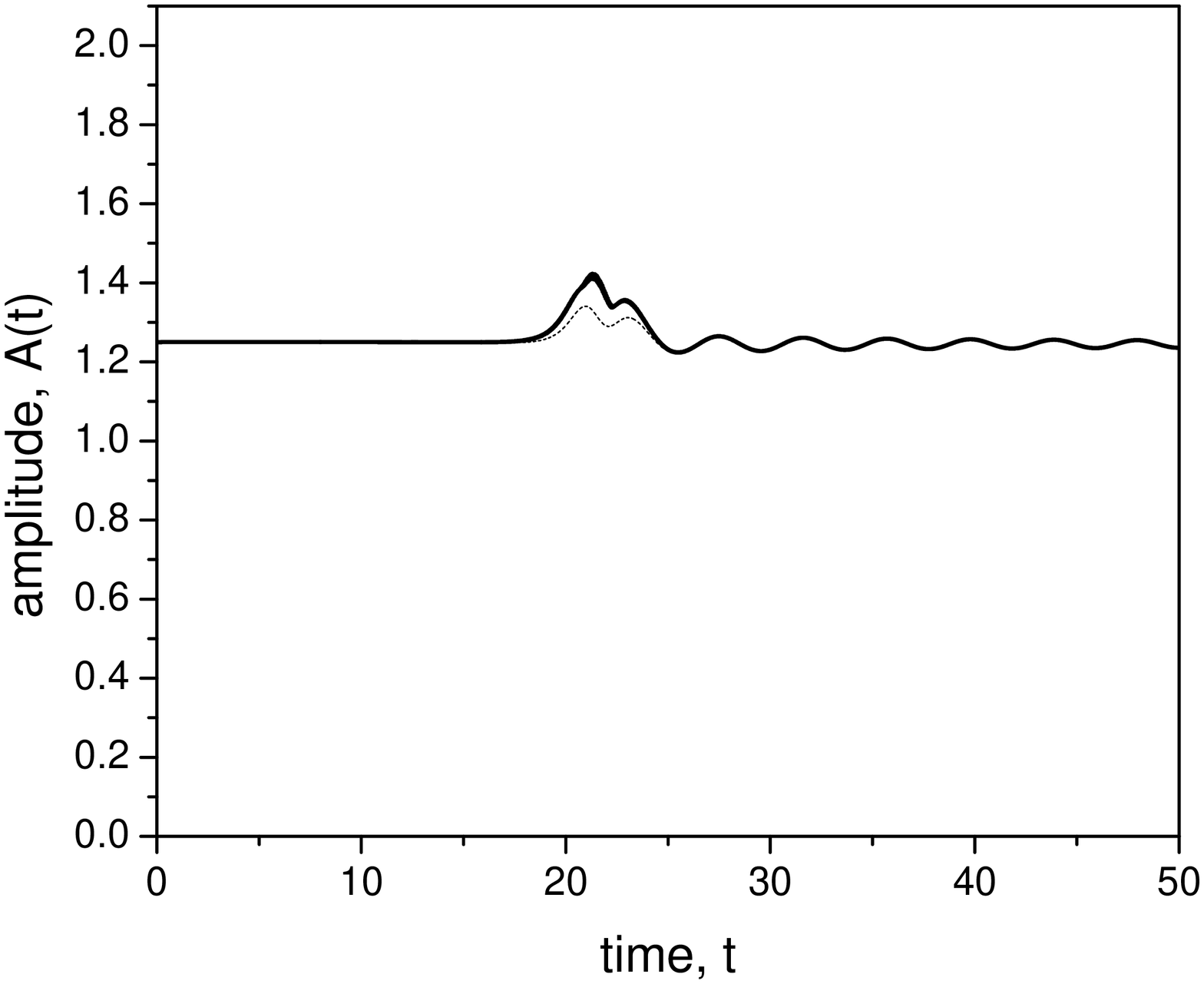}\quad
\includegraphics[width=6cm,height=6cm,angle=-90,clip]{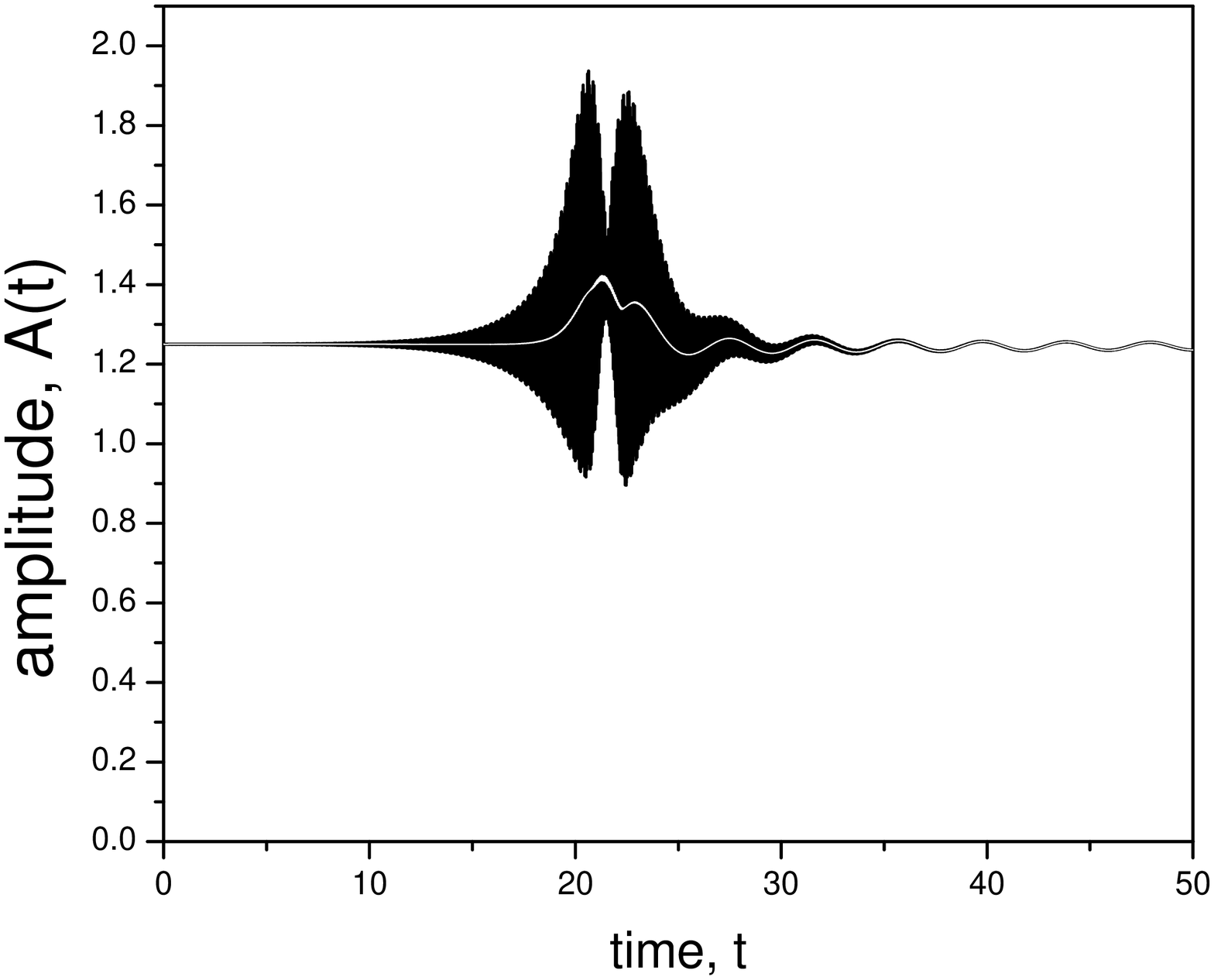}
\caption{Evolution of the NLS soliton amplitude: left panel shows
averaged amplitude of full equation (solid line) and amplitude of
averaged equation (dot line). Right panel depicts non-averaged
(black line)  and averaged (white line) amplitudes of full
equation.} \label{Fig2}
\end{figure}

First we studied travelling of solitons through imaginary
PT-potential Eq.~(\ref{PTpot}) (supposing $V_0$ to be zero). We
considered the NLS soliton having the form
\begin{equation}\label{Zahsol}
u_{sol}(x,t) = A\sech\left(A(x - vt)\right)\exp\left(i(vx/2 +(A^2
- v^2 /4)t) \right) .
\end{equation}
Fig. \ref{Fig1}  depicts dynamics of a NLS soliton being scattered
by this PT-potential. The soliton of the norm $N=2.5$ incident on
the PT-potential moves with the velocity $v=0.4$. The PT potential
Eq.~(\ref{PTpot}) with $V_0 = 0$ and $W_0 = 1.5$ is under action
of temporal modulation. Parameters of the modulation in full
equation (\ref{govGP}) are $ \alpha_0 = 0, \ \ \alpha_1 = 36$ and
$\Omega = 120$ that correspond to modulation parameter $\epsilon =
0.3$ in averaged equation (\ref{avGP}).  One can see that the
travelling soliton keeps ideally its form in the course of all its
movement. Averaged dynamics of the wave function amplitude is
depicted in Fig. \ref{Fig2}. Left panel shows that amplitude makes
a jump as the soliton passes the potential center. Solid line is
for {\it averaged solution} of full equation (\ref{govGP}), dot
line is for solution of {\it averaged equation} (\ref{avGP})
Non-averaged and averaged amplitudes of full equation
(\ref{govGP}) is shown in the right panel.

\begin{figure}[htb]
\includegraphics[width=6cm,height=6cm,angle=-90,clip]{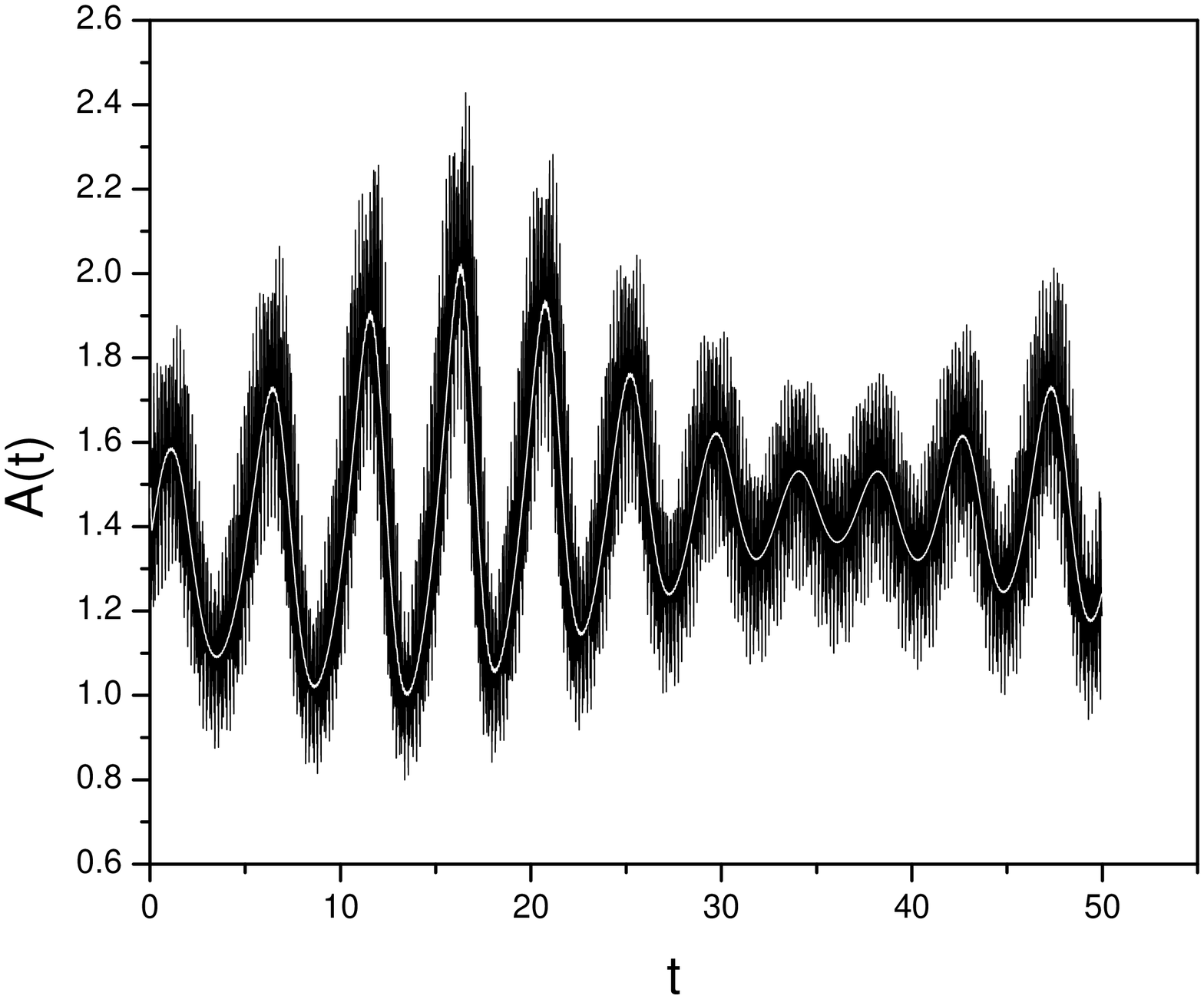}\quad
\includegraphics[width=6cm,height=6cm,angle=-90,clip]{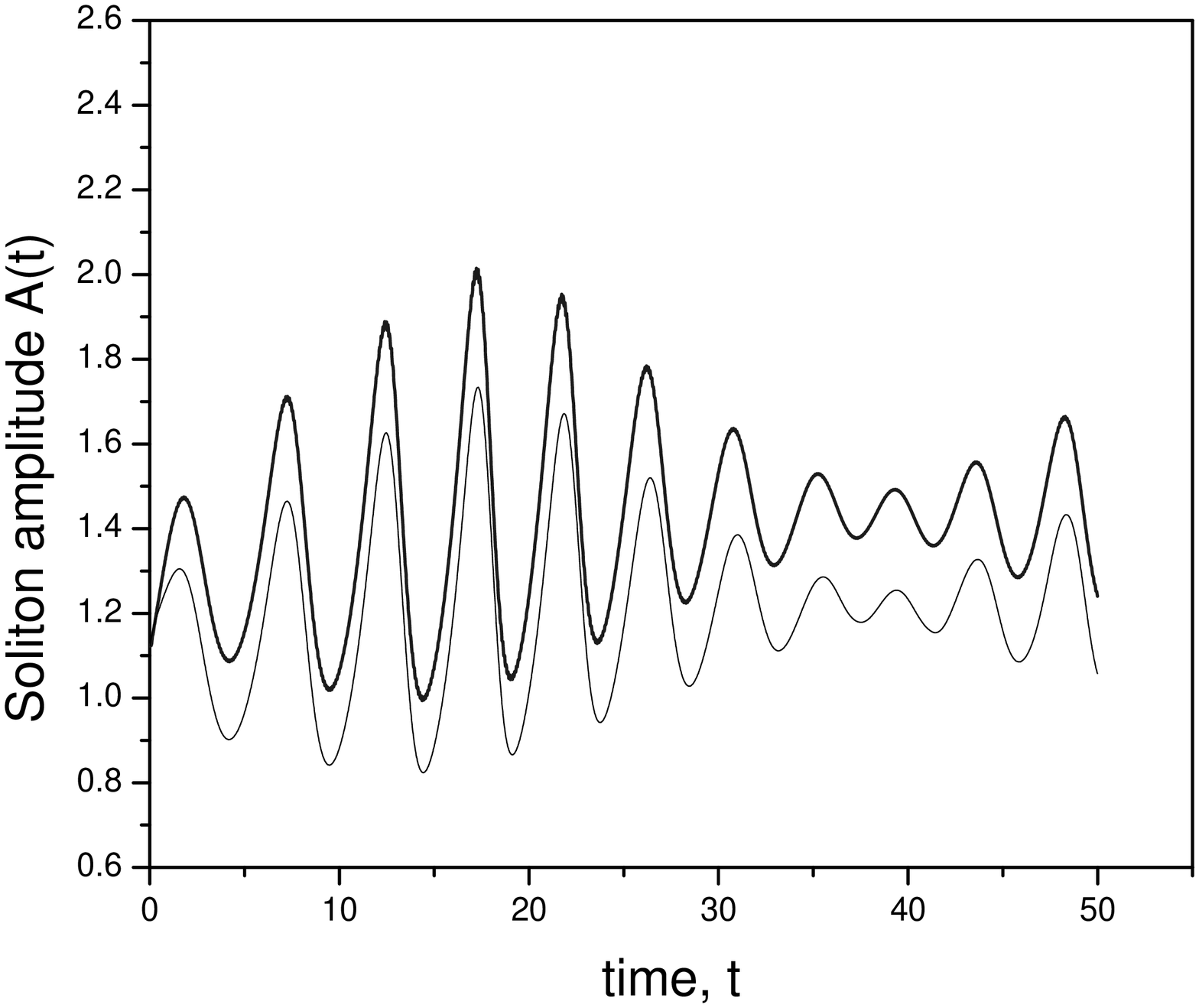}
\caption{Left panel: Time Dependence of the  amplitude of the full
equation solution (black line) and the averaged one (white line).
Right panel:  The dependence of the averaged amplitude of the full
equation solution on time (bold line) and amplitude of averaged
equation (dot line). PT soliton parameters are $V_0=1, W_0=1.5,
\epsilon =0.75$. } \label{amplitude}
\end{figure}

Next we studied averaged dynamics of the soliton solution
Eq.~(\ref{solMus}) under action of strong strength modulation of
PT-potential Eq.~(\ref{PTpot}). In the case of strong management
the results of full numerical simulations were compared with the
ones of simulation of the averaged equation Eq.~(\ref{lstavGP}).
In calculations $\alpha_0=1, \ \alpha_1=52.5, \Omega=70$ that
matches to $\epsilon=0.75$. Figure \ref{amplitude} depicts
oscillations of the soliton amplitude when initial position  of
the soliton is shifted and $x_0=-0.2$. The zero value $x_0=0$
corresponds to a stationary soliton solution. Parameters of the
potential Eq.~(\ref{PTpot}) $V_0 = 1, \ \ W_0 = 1.5$. In such a
choice of parameter values imaginary part $W_0$ is above the PT
threshold where a phase transition occurs and the spectrum enters
the complex domain i.e. the soliton solution becomes unstable.
Here two solutions are presented, {\it averaged solution} of the
equation Eq.~(\ref{govGP}) with rapidly varied strength of
potential Eq.~(\ref{PTpot}) and solution of the {\it averaged
equation} Eq.~(\ref{lstavGP}) with the effective potential .


\begin{figure}[htb]
\includegraphics[width=6cm,height=6cm,angle=-90,clip]{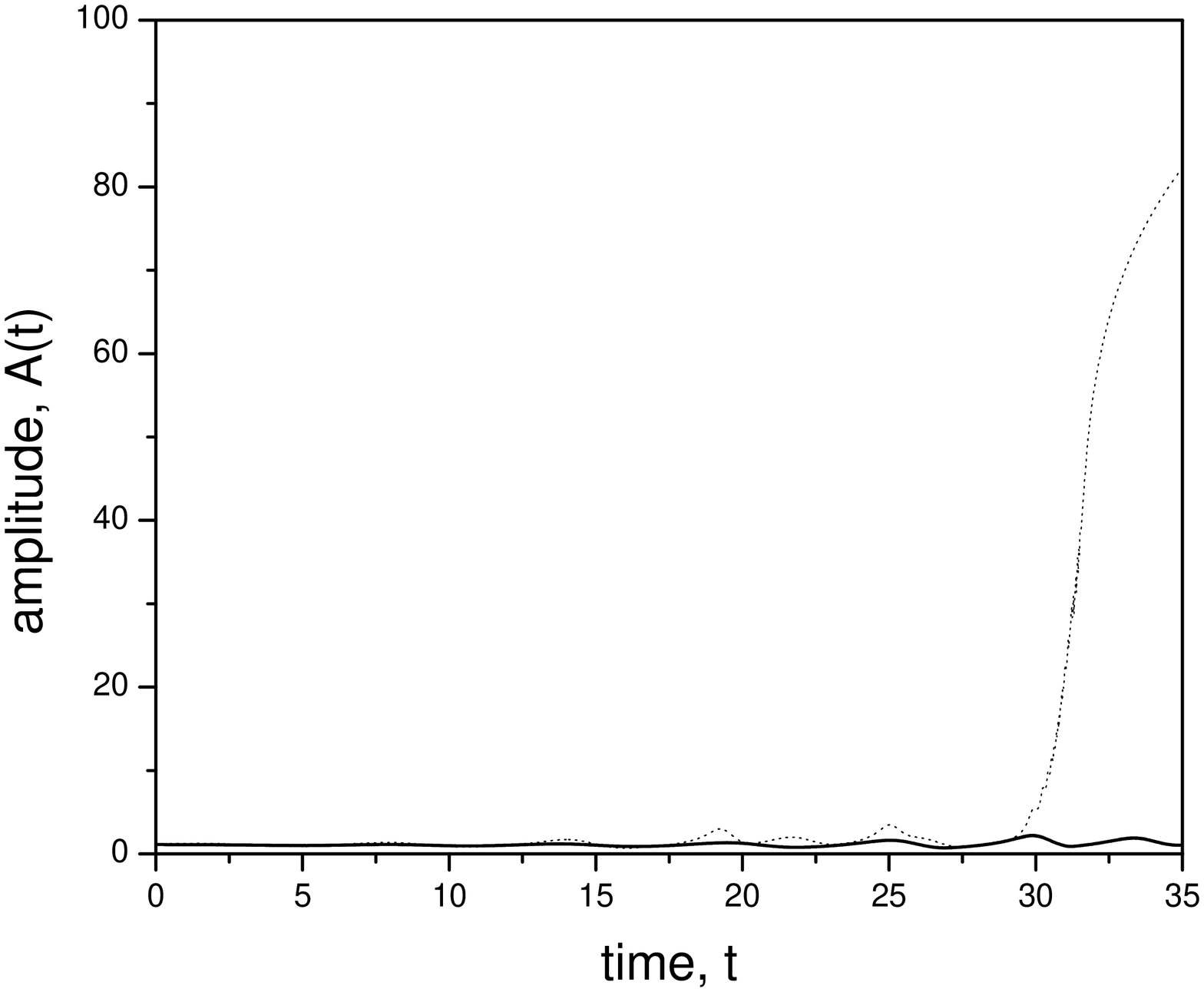}
\caption{The soliton amplitude evolution when rapid oscillations
of the potential is turned off, i.e. $\epsilon=0$ (solid line) and
evolution at $\epsilon=0.4$ (dot line), i.e. modulation is turned
on. PT potential parameters are $V_0=1, \ W_0=1.5, \ \alpha_0=1, \
\alpha_1=48, \ \Omega=120.$} \label{unstab}
\end{figure}

\begin{figure}[htb]
\includegraphics[width=6cm,height=6cm,angle=-90,clip]{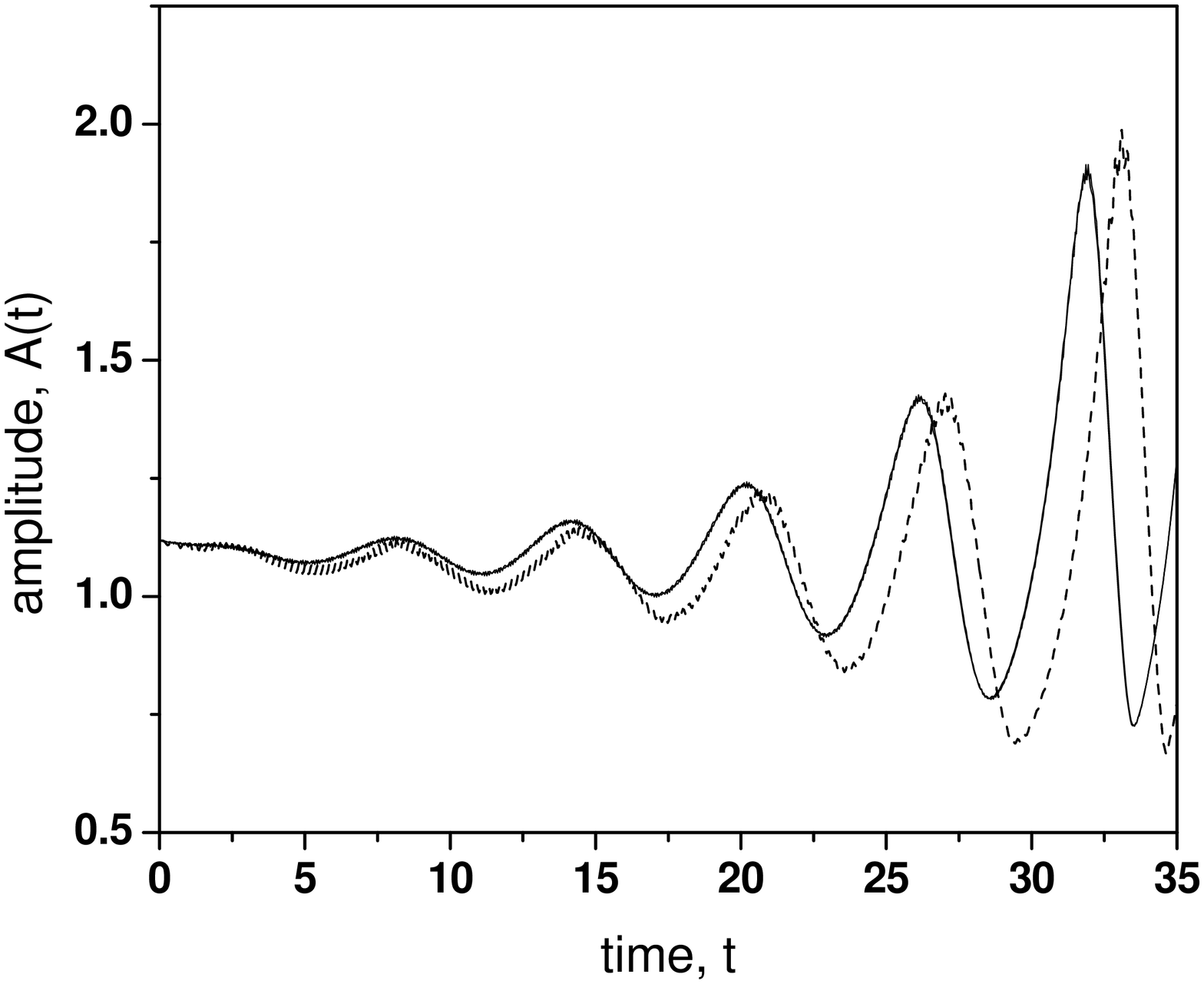}
\caption{The soliton amplitude evolutions are depicted at
$\epsilon=0.3$ with different rapid oscillations of the potential.
PT potential parameters are $V_0=1, \ W_0=1.5, \ \alpha_0=1, \
\alpha_1/\Omega = 9/30 \ ,  21/70 \ ,
 36/120$, respectively dash, solid, solid lines.} \label{conv}
\end{figure}

Figure \ref{unstab} shows the soliton amplitude evolutions when
rapid perturbations of the PT-potential Eq.~(\ref{PTpot}) are
turned off and so $\epsilon=0$ (solid line) and the case with
temporal modulations at $\epsilon=0.4$ (dot line). Parameters of
the potential Eq.~(\ref{PTpot}) are $V_0=1, \ \ W_0=1.5$. As seen
when $\epsilon=0$   the solution amplitude diverges ($W_0
> V_0 + 1/4$ in linear case) and the potential spectrum has complex domain.
Applying external modulation suppress the divergence.
Convergence of the parameter $\epsilon$ by the modulation
frequency variation is shown in Fig. \ref{conv}. Note that curves
for $\Omega = 70$ and $\Omega = 120$ coincide.


\section{Conclusion}
In conclusion we investigate the dynamics of optical solitons in
the media with rapidly varying PT-symmetric distribution of the
complex refractive index. We derive modified nonlinear Schrodinger
  equation to describe averaged dynamics of optical solitons.
Two cases with the weak and strong managements of
the complex refraction index  have been considered. In description
of the dynamics, the management results in appearance of effective
modified linear potential and spatially varying effective Kerr
nonlinearity in the NLSE, depending on the ratio of the amplitude
and the frequency of modulations. We study propagation of the NLS
soliton through oscillating PT-potential and found that the norm
of the soliton makes a jump when passing the potential, keeping
ideally its form. In the case of strong management $\epsilon \neq
0$ and when $W_0
> V_0 + 1/4$ (trap potential spectrum has complex domain) we
observe that strong management turns off the divergence of the
soliton dynamics being observed at $\epsilon = 0$,
{\it  i.e. we observe the dynamical stabilization phenomena}. It has been
theoretically and numerically shown that two parameters $\alpha_1$
and $\Omega$ form a single parameter $\epsilon =
\alpha_1/\Omega$.
These effects open new possibilities for the
control and steering of the optical beams.



\begin{thebibliography}{99}


\bibitem{KYZ}
V. V. Konotop, J. Yang, and D. A. Zezyulin,
Rev. Mod. Phys. {\bf 88}  (2016) 035002.

\bibitem{Dmitriev}
S. V. Suchkov, A. A. Sukhorukov, J. Huang, S. V. Dmitriev,
C. Lee, Yu. S. Kivshar,
Laser \&\ Photonics Reviews, 10 (2016) 177.

\bibitem{Bender}
C. M. Bender and S. Boettcher,
Phys. Rev. Lett. {\bf 80} (1998) 5243–5246.



\bibitem{Musslim}
Z.H.Musslimani, K.G.Makris, R. El-Ganainy, and D.N.
Christodoulides, Phys.Rev Letters {\bf 100} (2008) 030402.

\bibitem{AKKZ}
F. K. Abdullaev, Y. V. Kartashov, V. V. Konotop, and D. A.
Zezyulin,
Phys. Rev. A {\bf 83} (2011) 041805.

\bibitem{Kiv1}
X. B. Luo, J. H. Huang, H. H. Zhong, X. Z. Qin, Q. T. Xie,
Y. S. Kivshar, and C. H. Lee,
Phys. Rev. Lett. {\bf 110} (2013) 243902.

\bibitem{Mal1}
R. Driben and B. A. Malomed,
Europhys. Lett. {\bf 96} (2011) 51001.

\bibitem{Abd1}
R. L. Horne, J. Cuevas, P. G. Kevrekidis, N. Whitaker,
F. K. Abdullaev, and D. J. Frantzeskakis,
J. Phys. A {\bf 46} (2013) 485101–19.

\bibitem{Longhi}
B. Torosov, G.Della Valle, and S. Longhi,
Phys. Rev. A {\bf 88} (2013) 052106.

\bibitem{CW}
H. Cartarius and G. Wunner,  Phys. Rev. A 86 (2012) 013612.

\bibitem{AK}
 Yu. S. Kivshar, G.P. Agrawal, Optical Solitons: From Fibers To Photonic Crystals,
Academic Press, San Diego, 2003.


\bibitem{Oberthaler}
R. St\"utzle, M. C. G\"obel, Th. Hörner, E. Kierig, I. Mourachko, M. K. Oberthaler, M. A. Efremov, M. V. Fedorov, V. P. Yakovlev, K. A. H. van Leeuwen, and W. P. Schleich
Phys. Rev. Lett. {\bf 95} (2005) 110405.

\bibitem{Kivshar1995}
Y.S. Kivshar, K.H. Spatschek  Chaos, Solitons and Fractals {\bf
5}, (1995), 2551-2569.

\bibitem{AbdGal}
F.Kh. Abdullaev, R.M. Galimzyanov J.Phys B {\bf 36},(2003),
1099-1108.




\bibitem{Ahmed}
Z.  Ahmed, Phys. Lett.  A 282 (2001) 343-348.

\end{thebibliography}
\end{document}